# Estimating proton beam energy spread using Bragg peak measurement

V. Anferov[1], V. Derenchuk, R. Moore, A. Schreuder

ProNova Solutions LLC, Knoxville, TN.

## 1. Introduction

ProNova is installing and commissioning a two room proton therapy system in Knoxville, TN. Beam energy out of the 230 MeV cyclotron was measured on Jan 24, 2015. Cyclotron beam was delivered into a Zebra multi layered IC detector (IBA) calibrated in terms of penetration range in water. The analysis of the measured Bragg peak determines penetration range in water which can be subsequently converted into proton beam energy using Range-Energy tables [1]. We extended this analysis to obtain an estimate of the beam energy spread out of the cyclotron.

## 2. Materials and Methods

First we performed a set of Monte Carlo calculations of Bragg peak curves for monoenergetic proton beam with penetration range incrementing in steps of about 2cm up to maximum beam energy of 250MeV, which corresponds to about 38cm penetration range in water. We then analyzed the shape of the calculated Bragg peaks to define correlation between Bragg peak parameters (width at 50% and 80% dose levels, distal falloff) and the beam penetration range. The one-to-one correspondence can be readily explained. If we consider proton beam with energy E1 that exceeds reference energy E0, than it has to pass an additional slab of water until its energy is degraded down to the reference value E0. Passing through this additional slab, the proton beam acquires energy spread which changes the width of the resulting Bragg peak. In other words width characteristics of the Bragg peak define the energy of the monoenergetic proton beam E1 that would create such a Bragg peak shape.

Unfortunately, beam energy spread is not the only parameter defining the shape of the Bragg peak. It is well known that geometry factors such as field size and SAD or field divergence, impact the shape of Bragg peak. We then analyzed how this correlation changes when the shape of the Bragg peak is distorted by the beam focusing conditions. In other words, when instead of ideal large uniform field, the Bragg peak is measured in a large diverging beam spot. This analysis allows us to select the Bragg peak characteristic that is least sensitive to the beam focusing conditions.

The Bragg peak simulations were performed using MCNPx Monte Carlo code [2], where we setup several simple geometries. A large uniform field geometry, where a 10cm in diameter parallel beam source is setup in front of the water phantom containing a cylindrical mesh of detectors (2cm in diameter similar to that of Zebra detector) separated by 1mm in depth. A second geometry option was setup for small beam 2cm in diameter with adjustable divergence. The large field geometry option was used to establish Bragg peak shape parameter correlation with beam energy. The small diverging field option was used to study how Bragg peak shape varies with geometry changes. This option was also used to validate the beam energy spread estimates.

---

[1] Now at Varian Medical Systems Vladimir.anferov@varian.com

We then performed analysis of the experimentally measured Bragg peak. The parameters of the measured Bragg peak (width and distal fall off) correspond to penetration range R1 that is larger than observed R0. This suggests that the incoming beam had energy spread that can be estimated as energy spread that monoenergetic proton beam of energy E(R1) acquires after passing through the water slice with R1−R0 thickness.

### 3. Results

For large uniform field impinging on a small area detector, we observed linear dependence of each Bragg peak parameter on beam penetration range as shown in Figure 1.

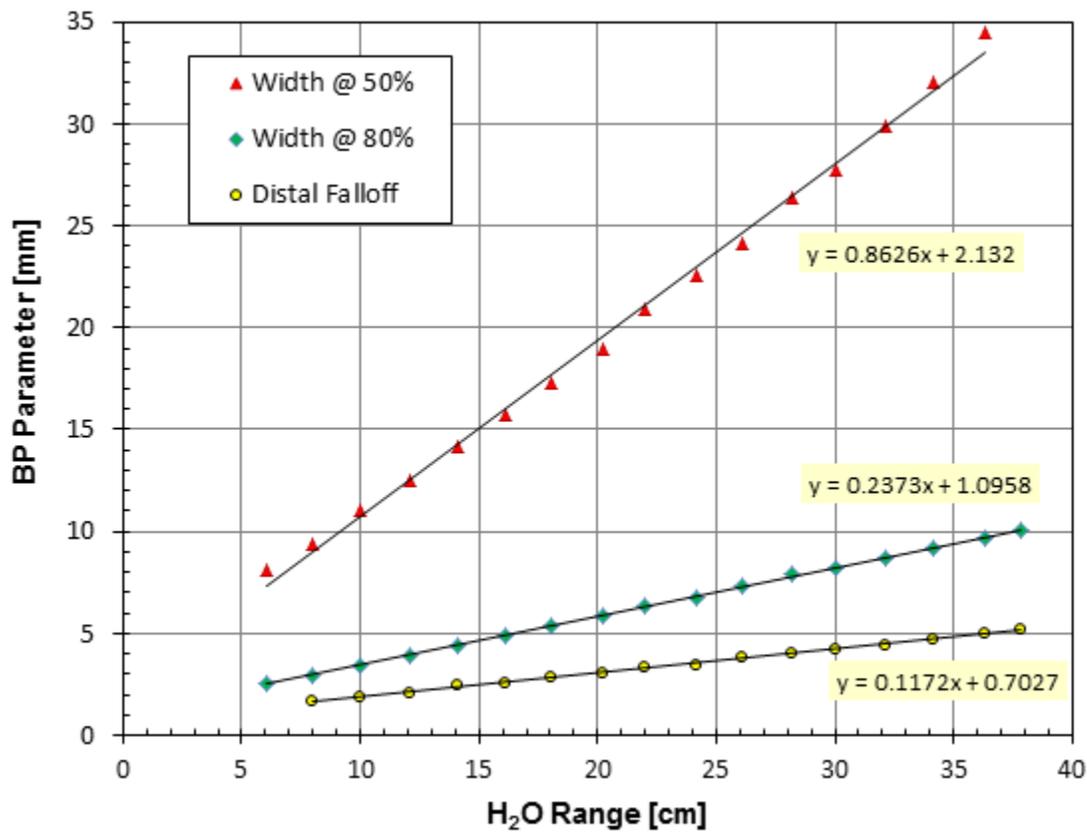

**Figure 1:** Correlation between penetration range of a monoenergetic proton beam and Bragg peak parameters.

Then we studied how this correlation changes when the shape of Bragg peak is distorted by the beam focusing conditions. This study was done at fixed beam energy of 182 MeV and variable beam source conditions. As illustrated in Figure 2, small field size or diverging beam cause Bragg peak deformation predominantly in the proximal region. The distal shape of the renormalized Bragg peaks stays nearly constant. This excludes usage of Bragg peak width parameters for energy spread estimates and we used distal falloff instead.

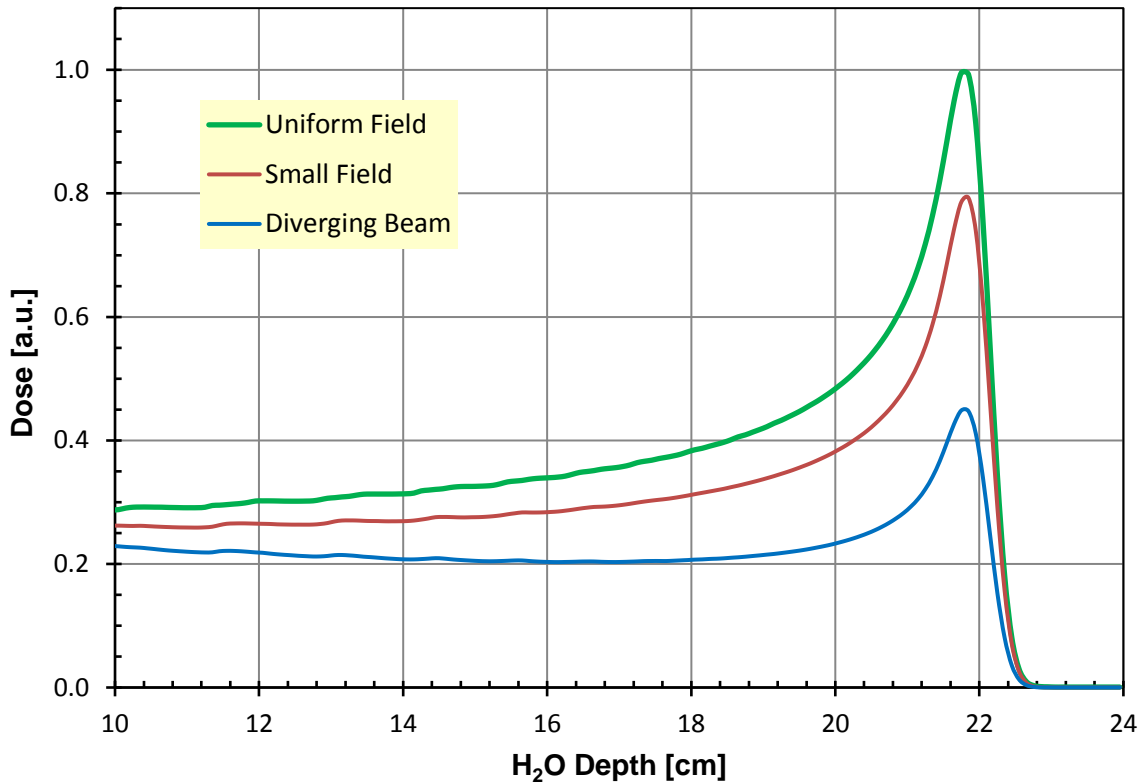

**Figure 2:** Changes in Bragg peak shape for different source geometry of a 182 MeV proton beam.

We then analyzed measured Bragg peak data to estimate the beam energy spread. The measured Bragg peaks had an average distal falloff of 4.86mm, which corresponds to an effective range of 35.5cm for a monoenergetic beam. The 32.7cm measured penetration range is 2.8cm less. Passage of a 230 MeV proton beam through a 2.8cm thick slab of water results in a ±0.56 MeV energy spread [3]. As a final check, we tested an agreement between shapes of the measured Bragg peak and one generated by Monte-Carlo code for proton beam with 0.56 MeV energy spread. Figure 3 shows the simulated Bragg peak curves for 230 MeV proton beam with 0.56 MeV energy spread, with measurement data points plotted on top of the simulated curves. Note the three simulated shapes of the Bragg peak were done at different divergence angles and the value of the cosine of the divergence angle is given in the graph legend. As discussed earlier, the changes in Bragg peak shape due to variable divergence occur in the proximal region only, while the distal shape stays constant.

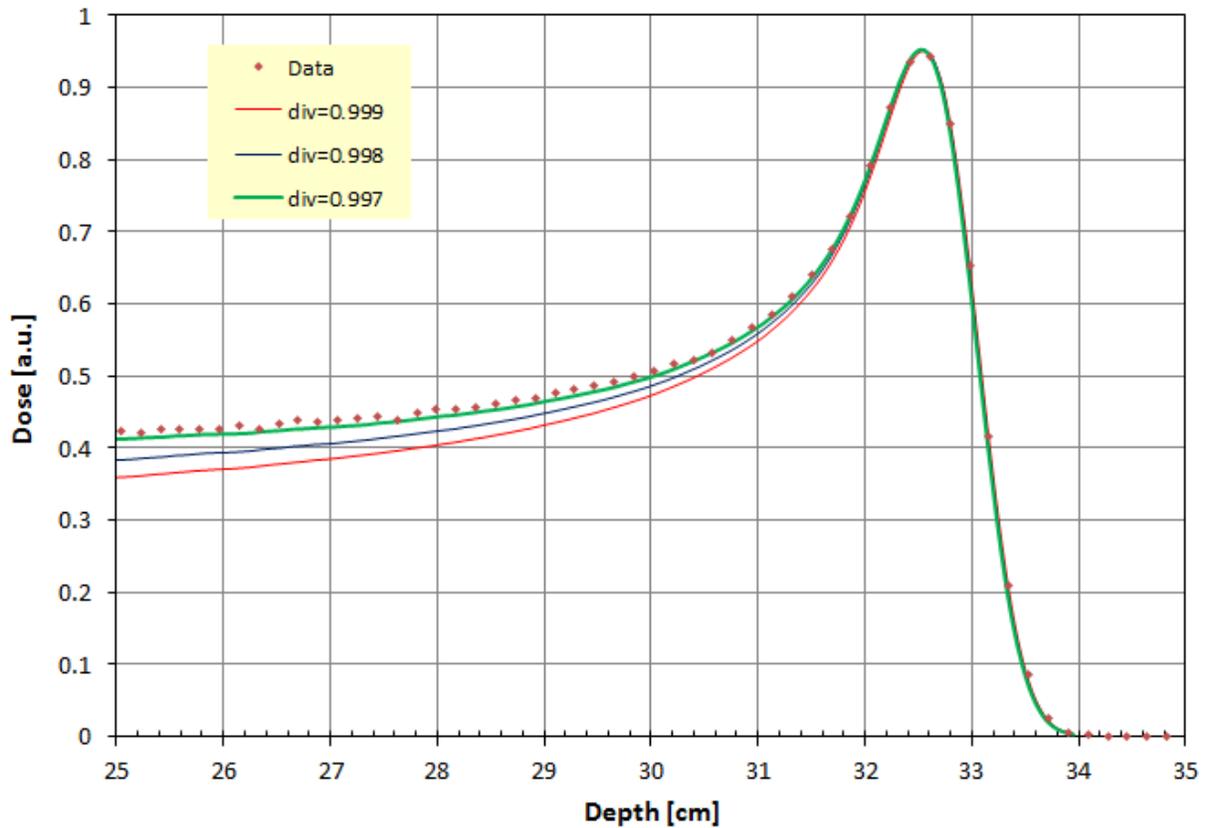

Figure 3: Measured Bragg peak data and simulated Bragg peaks for 231 MeV proton beam with 0.56 MeV energy spread.

## 4. Conclusions

We extended Bragg peak analysis to estimate the proton beam energy spread based upon the distal falloff value. Using Monte Carlo simulations we established the correlation between Bragg peak shape parameters (width at 50% and 80% dose levels, distal falloff) and penetration range for a monoenergetic proton beam. For large uniform field impinging on a small area detector, we observed linear dependence of each Bragg peak parameter on beam penetration range. Then we studied how this correlation changes when the shape of Bragg peak is distorted by the beam focusing conditions. As shown in Figure 2, small field size or diverging beam cause Bragg peak deformation predominantly in the proximal region; thus, we used distal falloff for beam energy spread estimates.

Using the measured Bragg peak data we obtained the range of the monoenergetic proton beam that would have the same distal falloff as one observed. The penetration range of this monoenergetic beam exceeds observed range by 2.8cm. Thus, the beam energy spread can be estimated by calculating the energy spread in 230MeV proton beam passing through a 2.8cm slab of water. As a final check, we confirmed agreement between shapes of the measured Bragg peak and one generated by Monte-Carlo code for proton beam with 0.56 MeV energy spread.

## 5. References

[1]. "Stopping Powers and Ranges for Protons and Alpha Particles", ICRU Report 49, (1993).

[2]. D.B. Pelowitz *et al*., "MCNPX User's Manual. Version 2.6.0", National Laboratory, Los Alamos (2008).

[3]. B. Gottschalk "BGWARE software tools for proton therapy",
http://huhepl.harvard.edu/~gottschalk